\documentclass{article}
\usepackage{arxiv}
\usepackage[T1]{fontenc}
\usepackage{microtype}
\usepackage{caption}
\usepackage{placeins}
\renewcommand{\headeright}{TCM-SEA}
\renewcommand{\undertitle}{}
\renewcommand{\shorttitle}{}

\newif\ifzh
\zhfalse

\usepackage{amsmath,amssymb}
\usepackage{graphicx}
\usepackage{booktabs}
\usepackage{multirow}
\usepackage[table]{xcolor}
\usepackage{tikz}
\usetikzlibrary{arrows.meta,positioning,fit,backgrounds,calc}
\usepackage[numbers,sort&compress]{natbib}
\usepackage{url}
\usepackage[hidelinks]{hyperref}

\definecolor{zhcolor}{HTML}{3A4A6B}
\definecolor{gain}{HTML}{2E7D32}
\definecolor{oursbg}{HTML}{EAF3EA}
\definecolor{cTerra}{HTML}{C0533A}
\definecolor{cNavy}{HTML}{2F3450}
\definecolor{cOchre}{HTML}{C0892B}
\definecolor{cSage}{HTML}{7BB59E}

\newcommand{\g}[1]{{\color{gain}\scriptsize(#1)}}

\newcommand{\bench}{TCM-MisDx}
\newcommand{\sys}{TCM-SEA}

\begin{document}

\title{Learning from Expert-Corrected Misdiagnoses:\\ A Self-Evolving Consultation Agent for Traditional Chinese Medicine after Failed Treatment%
}

\author{
Ye Wang$^{1}$ \quad
Zhikang Lin$^{2}$ \quad
Jia Liu$^{1}$ \quad
Tianle Zhang$^{1}$ \quad
Nanyue Wang$^{1,*}$ \\[0.9em]
\normalfont $^{1}$Experimental Research Center, China Academy of Chinese Medical Sciences, Beijing, China \\
\normalfont $^{2}$University of Science and Technology Beijing, Beijing, China
}
\date{}

\maketitle
{\renewcommand{\thefootnote}{*}\footnotetext{Corresponding author.}}

\begin{abstract}
In traditional Chinese medicine (TCM), treatment follows syndrome differentiation (\emph{bianzheng}), the step that maps a patient's findings to a TCM syndrome. A misdiagnosed syndrome therefore leads directly to inappropriate treatment, and the error usually surfaces only after that treatment fails. The physician who sees the patient next must diagnose a condition already altered by the wrong therapy, identify the initial physician's error, and avoid continuing the failed therapeutic direction. We introduce \bench{}, an interactive benchmark of 820 consultations after treatment failure built from three published monographs of TCM misdiagnosis cases. An agent gathers symptoms, tongue and pulse findings through the four diagnostic methods before reporting the current syndrome, the initial error, and a corrective treatment plan. We further present \sys{}, a self-evolving consultation agent that rewrites its own clinical skill document: failed consultations are compared with the expert physician's documented analysis, the agent proposes a minimal edit, and the edit is kept only if it passes an acceptance gate on held-in and held-out cases. The underlying language model is never updated. On the sealed test set, \sys{} raises the composite score from 61.8 to 65.5, syndrome accuracy from 53.9 to 57.2 and treatment-principle accuracy from 44.7 to 48.9, and lowers the therapeutic momentum rate, the fraction of corrective plans that keep the initial physician's failed direction, from 44.1 to 38.2. Learning from expert cases outperforms learning from outcome feedback alone, with the largest gains on the monograph of inappropriate-treatment cases. The two accepted edits are short, human-readable clinical checklists that physicians can audit directly.
\end{abstract}

\keywords{Traditional Chinese medicine \and syndrome differentiation \and misdiagnosis \and interactive diagnosis \and self-evolving agents \and LLM agents}

\section{Introduction}

Diagnostic error is among the most consequential threats to patient safety: an estimated 5.08\% of US adult outpatients experience one each year~\cite{singh2014}, and misdiagnosis-related serious harms reach about 795{,}000 disabilities or deaths annually~\cite{newmantoker2024}. In TCM, diagnosis takes the form of syndrome differentiation (\emph{bianzheng}), which maps the findings gathered through the four diagnostic methods (inspection, listening and smelling, inquiry, and palpation) to a TCM syndrome (\emph{zheng}); its reliability is low: three licensed practitioners examining the same 40 patients agreed in only 31.7\% of cases~\cite{zhang2005,obrien2009}. Because therapy follows directly from the differentiated syndrome, a misdiagnosis becomes an inappropriate treatment, and it is recognized only after the treatment fails.

Correcting such a case is not a matter of redoing the first encounter. The wrong therapy has already changed the illness, a situation classical TCM calls a deteriorated condition (\emph{huai bing}) and addresses with the maxim ``observe the pulse and symptoms, recognize which error was committed, and treat according to the present syndrome''~\cite{shanghanlun}. The correcting physician must diagnose the altered present state, name the initial physician's error, and choose a therapy that does not continue the failed direction. This setting is common in the documented record: in the three monographs we use, 58\% of cases record manifestations that appeared after the first treatment, and 29\% involve two or more successive inappropriate treatments.

Large language models (LLMs) now perform competitively on TCM licensing examinations and diagnostic benchmarks~\cite{liu2025tcm}, and domain systems inject knowledge through fine-tuning or retrieval~\cite{gastrotcm2026,dfglm2026,tcmdiffrag2026}. These evaluations present the complete case upfront and ask for a single answer; they do not test whether a model gathers what the initial physician overlooked, recognizes the prior error, or avoids repeating it. Across medicine, accuracy falls sharply once information must be acquired through dialogue~\cite{agentclinic2026}. Moreover, deployed systems are static: they do not improve from the errors they meet, and the knowledge they rely on cannot be inspected by clinicians.

TCM has, unusually, written this corrective expertise down. Misdiagnosis monographs record the failed treatment, the expert physician's analysis of what went wrong, and the corrective treatment that followed; clinicians learn to correct errors by studying exactly these cases, and consensus training on shared cases raises inter-practitioner agreement from 31.7\% to 73\%~\cite{zhang2008}. We turn this practice into an agent. Our contributions are:

\begin{itemize}
\item \textbf{\bench{}}, an interactive benchmark of 820 consultations after treatment failure with a deterministic patient simulator, hidden expert references, and a safety metric, the therapeutic momentum rate (TMR).
\item \textbf{\sys{}} (Self-Evolving Agent), an agent that rewrites its own clinical skill document from expert-corrected misdiagnosis cases under an acceptance gate, leaving the LLM weights untouched and producing edits clinicians can read.
\item \textbf{Evidence} that self-evolution improves accuracy and safety on a sealed test set, and that expert case analyses are a more useful learning signal than outcome feedback alone.
\end{itemize}

\section{Related Work}

\paragraph{LLMs for TCM.} Existing TCM systems specialize a model through domain corpora, knowledge graphs or retrieval~\cite{gastrotcm2026,dfglm2026,tcmdiffrag2026}, and are evaluated with static question answering~\cite{liu2025tcm}. Interactive TCM diagnosis has only recently been benchmarked~\cite{atcmd2026}, and no prior work targets the correction of a prior misdiagnosis.

\paragraph{Interactive and self-evolving clinical agents.} Sequential evaluation, where evidence is released only through actions, has become standard for clinical agents~\cite{agentclinic2026,medevoeval2026}, and recent systems accumulate diagnostic experience across cases~\cite{dxevolve2026}. In general-domain agents, verbal reflection~\cite{reflexion2023}, reflective prompt evolution~\cite{gepa2026}, evolving contexts~\cite{ace2026} and skill optimization~\cite{skillopt2026} improve an agent without weight updates. Self-Harness~\cite{selfharness2026} lets the same model diagnose its own failures and propose gated edits to its harness. We transfer this loop to TCM and replace generic failure evidence with the expert physician's documented analysis.

\section{The \bench{} Benchmark}

\paragraph{Source and episodes.} We digitized three published TCM misdiagnosis monographs: a collection of inappropriate-treatment cases (Source~A), a collection of corrected misdiagnoses (Source~B), and a collection of modern case records (Source~C). Each case was converted into an episode with three views following the clinical-episode format of MedEvoEval~\cite{medevoeval2026}. The \emph{patient view} holds the opening (demographics, chief complaint, and each prior treatment's diagnosis, principle, formula and response) and a list of atomic findings, each tagged with one of 12 inquiry categories and a temporal stage: present at the first visit, new after the prior medication, or elicited only on detailed questioning. The \emph{examination view} holds tongue, pulse, complexion and Western test results. The \emph{manager view}, hidden from the agent, holds the expert physician's syndrome, a rubric of the initial physician's errors, the corrective treatment plan normalized to a closed set of 24 treatment-principle elements, and the expert's original analysis.

\begin{table}[!htbp]
\centering
\caption{Statistics of \bench{}. }
\label{tab:data}
\small
\setlength{\tabcolsep}{4pt}
\begin{tabular}{lcccc}
\toprule
Source & Train & Val & Test & Total \\
\midrule
A: inappropriate treatment & 159 & 56 & 55 & 270 \\
B: corrected misdiagnosis & 56 & 18 & 14 & 88 \\
C: modern case records & 311 & 79 & 72 & 462 \\
\midrule
All & 526 & 153 & 141 & 820 \\
\midrule
\multicolumn{5}{l}{Findings per episode 5.08 \quad Exam items with content 2.35} \\
\multicolumn{5}{l}{Prior treatments per episode 1.37 \quad Error-rubric items 2.66} \\
\multicolumn{5}{l}{Episodes with post-medication findings 478 (58.3\%)} \\
\multicolumn{5}{l}{Episodes with $\geq$2 successive failed treatments 235 (28.7\%)} \\
\bottomrule
\end{tabular}
\end{table}

\paragraph{Interaction.} Each turn the agent takes one action: \texttt{ASK:}\emph{category}, \texttt{EXAMINE:}\emph{item}, or \texttt{FINALIZE} followed by a JSON answer with the present syndrome, the initial physician's errors, supporting evidence, the corrective treatment principle, ranked treatment-principle elements and a formula. The simulator is a deterministic lookup: it returns the recorded findings of the requested category with their temporal stage marker, or ``not recorded''. No LLM plays the patient, so every run sees identical information. The budget is 10 actions.

\paragraph{Metrics.} \emph{Syndrome accuracy}: the agent's syndrome matches the expert's in disease location and nature, judged by a fixed LLM judge (temperature 0). \emph{Treatment-principle accuracy}: the agent's first treatment-principle element equals the expert's primary element (exact match in the closed set). \emph{Misdiagnosis recognition}: the agent's stated prior errors hit at least one rubric item (same judge). \emph{TMR}: let $W$ be the elements used by the initial physician and $C$ the expert's corrective elements; the momentum set is $M = W\setminus(C\cup\{\text{other}\})$. For episodes with $M\neq\emptyset$ (130 of 141 test cases), TMR is the fraction in which the agent's elements intersect $M$, i.e.\ the corrective plan keeps a direction the expert abandoned. The composite averages the available parts:
\begin{equation}
S = \frac{1}{|P|}\sum_{p\in P} p, \quad P=\{\mathrm{Syn},\,\mathrm{Trt},\,\mathrm{Rec},\,1-\mathrm{TM}\},
\end{equation}
where the last term is included only when $M\neq\emptyset$.

\section{\sys{}: Self-Evolution from Expert-Corrected Cases}

\paragraph{Editable surface.} The agent's system prompt contains a fixed task description and a \emph{skill document}, a free-text section of clinical guidance. The skill document starts empty and is the only thing that changes; the LLM is frozen.

\paragraph{Evolution loop.} We follow the Self-Harness algorithm~\cite{selfharness2026} and implement it on the SkillOpt environment interface~\cite{skillopt2026}. Before the run, 80 training episodes are fixed as the held-in set by source-stratified sampling, and the 153 validation episodes serve as the held-out set. Each round: (1) the current skill document is run on both sets; (2) every failed held-in consultation is \emph{diagnosed} into a signature (terminal cause, criticality, agent mechanism), and signatures are clustered and ranked by severity and size into a brief, with successful cases listed as regression checks; (3) the agent \emph{proposes} one minimal edit (add, replace or delete a passage), and may decline if the evidence is weak; (4) the edited document is evaluated on both sets and \emph{accepted} only if the summed composite score does not drop on either set and rises on at least one. The same LLM performs consultation, diagnosis and proposal.

\paragraph{Learning from expert cases.} The distinctive signal in TCM is the expert physician's written analysis. In \sys{}, the diagnosis step of each failed case sees the per-metric outcome, the reference answers (syndrome, corrective treatment principle, error rubric) and the book's original expert analysis with the full corrective prescription. In the ablation \emph{w/o expert cases}, the diagnosis step sees only which metrics were right or wrong; all else is identical.

\section{Experiments}

\paragraph{Setup.} The consulting agent, the diagnosing and proposing agent, and the judge all use DeepSeek-Flash (\texttt{deepseek-flash}) with thinking disabled for the judge; the judge is called directly and kept fixed across all runs. Evolution runs for 9 rounds with one proposal per round and seed 42. The test set is sealed during evolution; each final skill document is evaluated on the 141 test episodes three times, and we report the mean.

\begin{table}[!htbp]
\centering
\caption{Results on the sealed test set (141 episodes, mean of three runs, \%). Relative change to the base harness in parentheses. }
\label{tab:main}
\resizebox{\textwidth}{!}{%
\setlength{\tabcolsep}{4pt}
\begin{tabular}{lcccccc}
\toprule
Method & Composite $\uparrow$ & Syndrome $\uparrow$ & Treat.\ principle $\uparrow$ & Recognition $\uparrow$ & TMR $\downarrow$ & Turns \\
\midrule
Base harness (empty skill document) & 61.8 & 53.9 & 44.7 & 93.1 & 44.1 & 9.12 \\
\sys{} after round 1 (1 edit) & 62.9 \g{+1.8\%} & 56.0 \g{+4.0\%} & 45.9 \g{+2.6\%} & 92.4 {\scriptsize($-$0.8\%)} & 42.3 \g{$-$4.1\%} & 9.12 \\
\textit{\sys{} w/o expert cases} & 64.8 \g{+4.9\%} & 56.0 \g{+4.0\%} & 47.5 \g{+6.4\%} & 94.8 \g{+1.8\%} & 38.5 \g{$-$12.8\%} & 9.00 \\
\rowcolor{oursbg}\textbf{\sys{}} (final, 2 edits) & \textbf{65.5} \g{+6.1\%} & \textbf{57.2} \g{+6.1\%} & \textbf{48.9} \g{+9.5\%} & \textbf{94.8} \g{+1.8\%} & \textbf{38.2} \g{$-$13.4\%} & 9.13 \\
\bottomrule
\end{tabular}}
\end{table}

\begin{figure}[!htbp]
\centering
\begin{minipage}[t]{0.49\textwidth}\centering
\includegraphics[width=\linewidth]{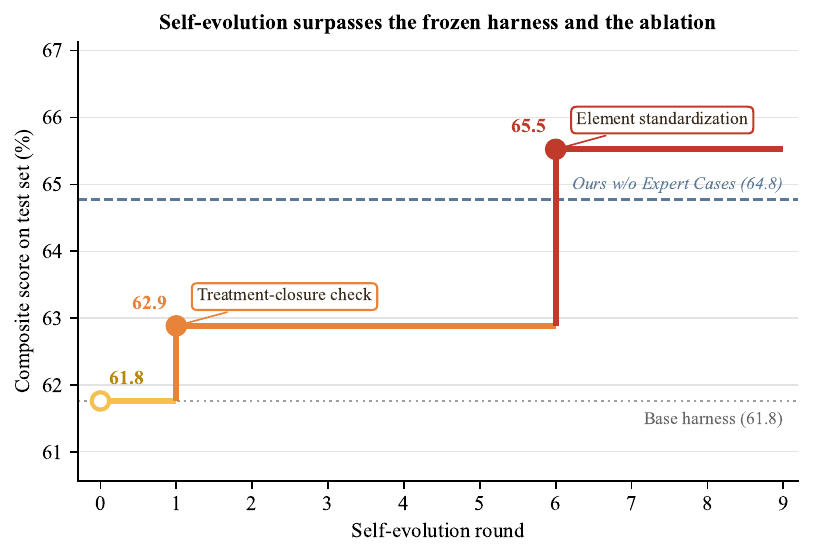}\\[-2pt]
{\small (a)}
\end{minipage}\hfill
\begin{minipage}[t]{0.49\textwidth}\centering
\includegraphics[width=\linewidth]{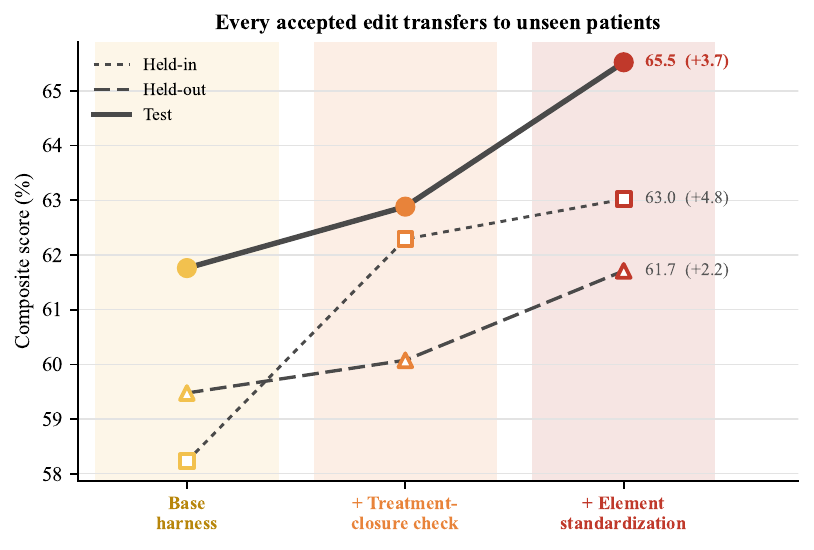}\\[-2pt]
{\small (b)}
\end{minipage}
\caption{Self-evolution of \sys{}. (a) Test-set composite score over the 9 rounds; each step is an accepted edit, and the final score exceeds both the frozen base harness and the ablation without expert cases. (b) Composite score on the held-in, held-out and sealed test sets after each accepted edit; every edit selected by the gate also raises the score on unseen test patients.
}
\label{fig:evo}
\end{figure}

\paragraph{Main results.} Table~\ref{tab:main} and Fig.~\ref{fig:evo}a summarize the sealed test results. Starting from an empty skill document, \sys{} raises the composite score from 61.8 to 65.5 (+3.7 points), syndrome accuracy from 53.9 to 57.2, and treatment-principle accuracy from 44.7 to 48.9, and lowers TMR from 44.1 to 38.2. In headroom terms, the final document closes 9.8\% of the composite gap to a perfect score, 24.1\% for misdiagnosis recognition and 13.4\% for TMR. The number of consultation turns is unchanged (9.12 vs.\ 9.13), so the gains come from better use of the same information rather than longer questioning. Improvement reaches every case source (Table~\ref{tab:source}): Source~A 57.5$\to$61.6, Source~B 53.6$\to$60.7, Source~C 66.6$\to$69.5.

\paragraph{The gate selects edits that transfer.} Over 9 rounds, two edits were accepted, at rounds 1 and 6. Each accepted edit raised the score on the held-in set (58.2$\to$62.3$\to$63.0), on the held-out set (59.5$\to$60.1$\to$61.7), and on the sealed test set (61.8$\to$62.9$\to$65.5; Fig.~\ref{fig:evo}b). The test set was never seen during selection, so the gate's criterion on held-in and held-out cases carried over to unseen patients.

\begin{figure}[!htbp]
\centering
\includegraphics[width=\textwidth]{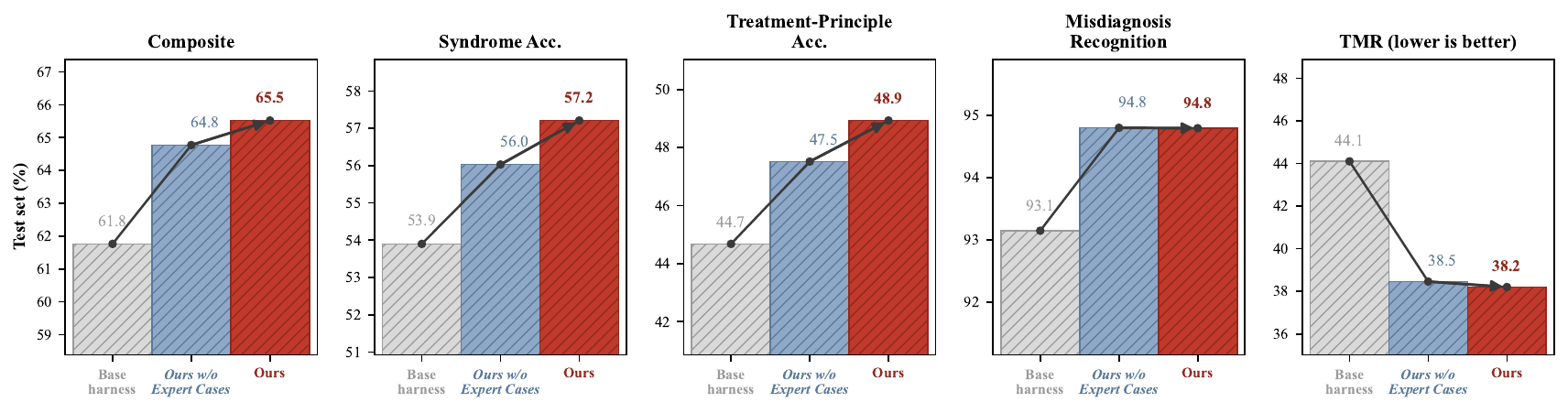}
\caption{Ablation on the learning signal (test set, \%). Base harness, \sys{} learning from outcome feedback only (w/o expert cases), and \sys{} learning from expert cases.
}
\label{fig:abl}
\end{figure}

\begin{table}[!htbp]
\centering
\caption{Test results per case source (\%). Base / w/o expert cases / \sys{}. }
\label{tab:source}
\footnotesize
\setlength{\tabcolsep}{8pt}
\begin{tabular}{lccc}
\toprule
Source ($n$) & Composite & Treat.\ principle & TMR $\downarrow$ \\
\midrule
A (55) & 57.5 / 60.6 / \textbf{61.6} & 37.0 / 40.6 / \textbf{46.7} & 47.6 / 43.5 / 44.2 \\
B (14) & 53.6 / 58.9 / \textbf{60.7} & 14.3 / 19.0 / \textbf{21.4} & 28.6 / 23.8 / \textbf{19.0} \\
C (72) & 66.6 / 69.1 / \textbf{69.5} & 56.5 / 58.3 / 56.0 & 44.8 / 37.8 / 37.8 \\
\bottomrule
\end{tabular}
\end{table}

\paragraph{Expert cases are the better teacher.} With identical loop, budget and seed, learning from expert cases outperforms learning from outcome feedback on composite (65.5 vs.\ 64.8), syndrome (57.2 vs.\ 56.0), treatment principle (48.9 vs.\ 47.5) and TMR (38.2 vs.\ 38.5), with equal recognition (Fig.~\ref{fig:abl}). The advantage is largest where the expert text is most informative about therapy: on Source~A, the inappropriate-treatment monograph, treatment-principle accuracy reaches 46.7 with expert cases versus 40.6 without and 37.0 at baseline; on Source~B, \sys{} attains the lowest TMR, 19.0 (Table~\ref{tab:source}).

\begin{table}[!htbp]
\centering
\caption{Accepted skill-document edits (condensed translation; originals in Chinese). }
\label{tab:edits}
\small
\renewcommand{\arraystretch}{1.1}
\begin{tabular}{p{0.95\textwidth}}
\toprule
\textbf{\sys{}, R1: Treatment-closure check.} Before finalizing, list the pathogenic keywords of the syndrome (deficiency, cold, heat, phlegm, stasis, stagnation, \ldots) and confirm each has a matching treatment element; if the initial treatment direction has been falsified (e.g.\ bitter-cold damaged yang), do not keep that direction as the primary element. \\

\midrule
\textbf{\sys{}, R6: Element standardization.} Map each element to standard TCM terms, remove elements unrelated to the core treatment principle (e.g.\ draining water, astringing, nourishing yin unless explicitly indicated), and correct omissions or redundancies. \\

\midrule
\textbf{w/o expert cases, R2: Error-direction self-audit.} Before finalizing, state the initial physician's wrong direction and its core elements, check whether the current elements and formula still contain them, and replace them with the corrective direction. \\

\bottomrule
\end{tabular}
\end{table}

\paragraph{What the agent learned.} Table~\ref{tab:edits} lists every accepted edit. All are short pre-submission checklists written in clinical language: \sys{} learned to close the loop between the differentiated pathogenesis and the chosen treatment elements, and to discard elements that do not serve the core principle; the outcome-only variant learned a single audit against the initial treatment direction. Because the entire learned state is this document, a physician can read, approve, or strike any line before deployment.

\section{Conclusion}

We framed a clinically central but unbenchmarked TCM task, consultation after failed treatment, as an interactive benchmark with a safety metric for therapeutic momentum, and showed that an agent can improve at it by rewriting its own clinical skill document from expert-corrected misdiagnosis cases while its language model stays frozen. On a sealed test set, \sys{} improves syndrome and treatment-principle accuracy and reduces the continuation of failed therapy, and expert case analyses provide a stronger learning signal than outcome feedback. The learned knowledge takes the form of short, auditable clinical checklists, which mirrors how case-based expertise is reviewed and passed on in TCM practice. Future work will add evaluation on further LLMs and physician assessment of the learned guidance.

\bibliographystyle{unsrtnat}
\bibliography{refs}

@article{singh2014,
  author  = {Singh, Hardeep and Meyer, Ashley N. D. and Thomas, Eric J.},
  title   = {The frequency of diagnostic errors in outpatient care: estimations from three large observational studies involving {US} adult populations},
  journal = {BMJ Quality \& Safety},
  year    = {2014},
  volume  = {23},
  number  = {9},
  pages   = {727--731}
}

@article{newmantoker2024,
  author  = {Newman-Toker, David E. and others},
  title   = {Burden of serious harms from diagnostic error in the {USA}},
  journal = {BMJ Quality \& Safety},
  year    = {2024},
  volume  = {33},
  number  = {2},
  pages   = {109--120}
}

@article{zhang2005,
  author  = {Zhang, Grant G. and others},
  title   = {Variability in the traditional {Chinese} medicine ({TCM}) diagnoses and herbal prescriptions provided by three {TCM} practitioners for 40 patients with rheumatoid arthritis},
  journal = {Journal of Alternative and Complementary Medicine},
  year    = {2005},
  volume  = {11},
  number  = {3},
  pages   = {415--421}
}

@article{zhang2008,
  author  = {Zhang, Grant G. and others},
  title   = {Enhancing reliability and validity of syndrome differentiation in traditional {Chinese} medicine through consensus training},
  journal = {Journal of Alternative and Complementary Medicine},
  year    = {2008},
  volume  = {14},
  number  = {5},
  pages   = {503--508}
}

@article{obrien2009,
  author  = {O'Brien, Kylie A. and Birch, Stephen},
  title   = {A review of the reliability of traditional {East Asian} medicine diagnoses},
  journal = {Journal of Alternative and Complementary Medicine},
  year    = {2009},
  volume  = {15},
  number  = {4},
  pages   = {353--366}
}

@article{liu2025tcm,
  author  = {Liu and others},
  title   = {Evaluating the role of large language models in traditional {Chinese} medicine diagnosis and treatment recommendations},
  journal = {npj Digital Medicine},
  year    = {2025},
  doi     = {10.1038/s41746-025-01845-2}
}

@article{gastrotcm2026,
  author  = {{GastroTCM Team}},
  title   = {{GastroTCM}: a large language model assistant for gastroenterology in traditional {Chinese} medicine},
  journal = {Chinese Medicine},
  year    = {2026},
  doi     = {10.1186/s13020-025-01295-8}
}

@article{dfglm2026,
  author  = {{DFGLM-TCM Team}},
  title   = {{DFGLM-TCM}: an integrated knowledge- and experience-driven large language model system for traditional {Chinese} medicine practice},
  journal = {Chinese Medicine},
  year    = {2026},
  doi     = {10.1186/s13020-026-01512-y}
}

@article{tcmdiffrag2026,
  author  = {{TCM-DiffRAG Team}},
  title   = {{TCM-DiffRAG}: personalized syndrome differentiation reasoning method for traditional {Chinese} medicine based on knowledge graph and chain of thought},
  journal = {Frontiers in Medicine},
  year    = {2026},
  doi     = {10.3389/fmed.2026.1804478}
}

@article{atcmd2026,
  author  = {{ATCMD-Bench Team}},
  title   = {{ATCMD-Bench}: benchmarking agentic traditional {Chinese} medicine diagnosis},
  journal = {Pattern Recognition},
  year    = {2026},
  volume  = {179},
  pages   = {113679}
}

@article{agentclinic2026,
  author  = {Schmidgall, Samuel and others},
  title   = {{AgentClinic}: a multimodal benchmark for tool-using clinical {AI} agents},
  journal = {npj Digital Medicine},
  year    = {2026},
  doi     = {10.1038/s41746-026-02674-7}
}

@article{medevoeval2026,
  author  = {Zhang, Hui and others},
  title   = {{MedEvoEval}: Evaluating continual evolution of doctor agents through simulated clinical episodes},
  journal = {arXiv preprint arXiv:2606.28900},
  year    = {2026}
}

@article{dxevolve2026,
  author  = {{DxEvolve Team}},
  title   = {{DxEvolve}: self-evolving deep clinical research with interactive diagnosis},
  journal = {arXiv preprint arXiv:2603.10677},
  year    = {2026}
}

@inproceedings{reflexion2023,
  author    = {Shinn, Noah and Cassano, Federico and Gopinath, Ashwin and Narasimhan, Karthik and Yao, Shunyu},
  title     = {Reflexion: Language agents with verbal reinforcement learning},
  booktitle = {Advances in Neural Information Processing Systems},
  year      = {2023}
}

@inproceedings{gepa2026,
  author    = {Agrawal, Lakshya A. and Tan, Shangyin and Soylu, Dilara and Ziems, Noah and Khare, Rishi and Opsahl-Ong, Krista and Singhvi, Arnav and Shandilya, Herumb and Ryan, Michael J. and Jiang, Meng and Potts, Christopher and Sen, Koushik and others},
  title     = {{GEPA}: Reflective prompt evolution can outperform reinforcement learning},
  booktitle = {International Conference on Learning Representations},
  year      = {2026}
}

@inproceedings{ace2026,
  author    = {Zhang, Qizheng and others},
  title     = {Agentic context engineering: Evolving contexts for self-improving language models},
  booktitle = {International Conference on Learning Representations},
  year      = {2026}
}

@article{skillopt2026,
  author  = {Yang, Yifan and Gong, Ziyang and Huang, Weiquan and Yang, Qihao and Zhou, Ziwei and Huang, Zisu and Li, Yan and Gao, Xuemei and Dai, Qi and Liu, Bei and others},
  title   = {{SkillOpt}: Executive strategy for self-evolving agent skills},
  journal = {arXiv preprint arXiv:2605.23904},
  year    = {2026}
}

@article{selfharness2026,
  author  = {Zhang, Hangfan and Zhang, Shao and Li, Kangcong and Zhang, Chen and Chen, Yang and Zhang, Yiqun and Bai, Lei and Hu, Shuyue},
  title   = {{Self-Harness}: Harnesses that improve themselves},
  journal = {arXiv preprint arXiv:2606.09498},
  year    = {2026}
}

@book{shanghanlun,
  author    = {Zhang, Zhongjing},
  title     = {Treatise on Cold Damage ({Shang Han Lun})},
  note      = {Clause 16},
  year      = {c. 200 CE}
}

\end{document}